\documentclass[prb,reprint,a4paper,floatfix,superscriptaddress]{revtex4-1}

\usepackage{graphicx}
\usepackage{amsmath}
\usepackage{amssymb}
\usepackage{hyperref}
\usepackage{url}

\begin{document}

\title{Granular superconductivity and magnetic-field-driven recovery of macroscopic coherence in a cuprate/manganite multilayer }

\author{B.P.P. Mallett}
\email{benjamin.mallett@gmail.com}
\affiliation{University of Fribourg, Department of Physics and Fribourg Center for Nanomaterials, Chemin du Musée 3, CH-1700 Fribourg, Switzerland}

\author{J. Khmaladze}
\email{jarji.khmaladze@unifr.ch}
\affiliation{University of Fribourg, Department of Physics and Fribourg Center for Nanomaterials, Chemin du Musée 3, CH-1700 Fribourg, Switzerland}

\author{P. Marsik}
\email{premysl.marsik@unifr.ch}
\affiliation{University of Fribourg, Department of Physics and Fribourg Center for Nanomaterials, Chemin du Musée 3, CH-1700 Fribourg, Switzerland}

\author{E. Perret}
\affiliation{University of Fribourg, Department of Physics and Fribourg Center for Nanomaterials, Chemin du Musée 3, CH-1700 Fribourg, Switzerland}

\author{A. Cerreta}
\affiliation{University of Fribourg, Department of Physics and Fribourg Center for Nanomaterials, Chemin du Musée 3, CH-1700 Fribourg, Switzerland}

\author{M. Orlita}
\affiliation{Laboratoire National des Champs Magnétiques Intenses, CNRS-UJF-UPS-INSA, avenue des Martyrs, 38042 Grenoble, France}

\author{N. Bi\v{s}kup}
\affiliation{Departamento de Fısica Aplicada III, Instituto Pluridisciplinar, Universidad Complutense de Madrid, Spain
}

\author{M. Varela}
\affiliation{Departamento de Fısica Aplicada III, Instituto Pluridisciplinar, Universidad Complutense de Madrid, Spain
}

\author{C. Bernhard}
\email{christian.bernhard@unifr.ch}
\affiliation{University of Fribourg, Department of Physics and Fribourg Center for Nanomaterials, Chemin du Musée 3, CH-1700 Fribourg, Switzerland}

\date{\today}

\pacs{ }

\begin{abstract}

We show that in Pr$ _{0.5} $La$ _{0.2} $Ca$ _{0.3} $MnO$ _{3} $/YBa$ _{2} $Cu$ _{3} $O$ _{7} $ (PLCMO/YBCO) multilayers the low temperature state of YBCO is very resistive and resembles the one of a granular superconductor or a frustrated Josephson-junction network. Notably, a coherent superconducting response can be restored with a large magnetic field which also suppresses the charge-orbital order in PLCMO. This coincidence suggests that the granular superconducting state of YBCO is induced by the charge-orbital order of PLCMO. The coupling mechanism and the nature of the induced inhomogeneous state in YBCO remain to be understood.

\end{abstract}

\maketitle

The quantum phase transition from a superconducting (SC) to a metallic or even insulating state is of great scientific and technological interest \cite{gantmakher2010, goldman2010}. It is typically induced with disorder, electronic doping, or by decreasing the layer thickness of thin films. Especially interesting are the cases for which the transition can be controlled with an external field, as in electric-field effect devices of LaAlO$ _{3} $/SrTiO$ _{3} $ \cite{caviglia2008} and La$ _{2-x} $Sr$ _{x} $CuO$ _{4} $ \cite{bollinger2011}. Magnetic fields can also induce such a quantum phase transition, since they generally reduce the SC phase coherence. 
Only a few examples are known where coherent SC is restored by a magnetic field. In the bulk, these include the chevrel phase Eu$ _{x} $Sn$ _{1+x} $Mo$ _{6} $S$ _{8} $ \cite{meul1984} and the organic $ \lambda $-(BETS)$ _{2} $FeCl$ _{4} $ \cite{uji2001,balicas2001} for which the magnetic field compensates a negative exchange field from magnetic ions, thereby reducing the pair-breaking (Jacarino-Peter effect) \cite{jaccarino1962}, or suppresses detrimental magnetic fluctuations \cite{maekawa1978}. A reentrance of SC was also reported in Zn nanowires where the field seems to reduce quantum fluctuations by generating dissipative quasi-particles \cite{tian2005, chen2009}. 

In the following, we show that yet another kind of these rare cases can be found in cuprate/manganite multilayers. Here the magnetic field restores a coherent SC state in a thin YBa$ _{2} $Cu$ _{3} $O$ _{7} $  (YBCO) layer, most likely since it suppresses an interaction with the neighboring Pr$ _{0.5} $La$ _{0.2} $Ca$ _{0.3} $MnO$ _{3} $ (PLCMO) layers that is detrimental to a macroscopic SC coherence. PLCMO exhibits a combined charge-orbital order \cite{tokura1999} and an antiferromagnetic (AF) order with a weak ferromagnetic (FM) component that arises either from a spin canting or phase segregation. A large magnetic field suppresses this charge-orbital ordered AF state towards an itinerant FM state \cite{tokura1999}. Surprisingly, the latter state of PLCMO is less detrimental to the SC in YBCO than the former. 

PLCMO(20 nm)/YBCO(7, 9 and 20 nm)/PLCMO(20 nm) trilayers and PLCMO(20 nm)-YBCO(3.5, 4.5 nm) bilayers were grown on La$ _{0.3} $Sr$ _{0.7} $Al$ _{0.65} $Ta$ _{0.35} $O$ _{3} $ (LSAT) substrates by pulsed laser deposition. Details of the growth and characterization can be found in Refs.~\onlinecite{pypPRLsom, malik2012}. The (magneto-)resistance and vibrating sample magnetometry (VSM) measurements were made in a Quantum Design PPMS as described in Ref.~\onlinecite{pypPRLsom}. The optical response was determined with spectroscopic ellipsometry. In the terahertz (THz) region ($ 3$-$70 $~cm$ ^{-1} $) we used a home-built time-domain THz ellipsometer \cite{matsumoto2011, marsik2016}, in the infrared ($ 70$-$4000 $~cm$ ^{-1} $) a home-build setup attached to a Bruker 113v Fourier-transform spectrometer (FTIR) \cite{bernhard2004thinfilms} and, in the near-infrared to ultraviolet ($ 4000$-$52000 $~cm$ ^{-1} $), a Woollam VASE ellipsometer. Optical transmission at $ 4.2 $~K in magnetic fields up to 11~Tesla were measured at the Laboratoire National des Champs Magn\'{e}tiques Intenses (LNCMI) with a Bruker 113v FTIR for the THz and a 66v for the MIR-NIR ranges. Further details are given in Ref.~\onlinecite{pypPRLsom}.

\begin{figure*}
		\includegraphics[width=0.85\textwidth]{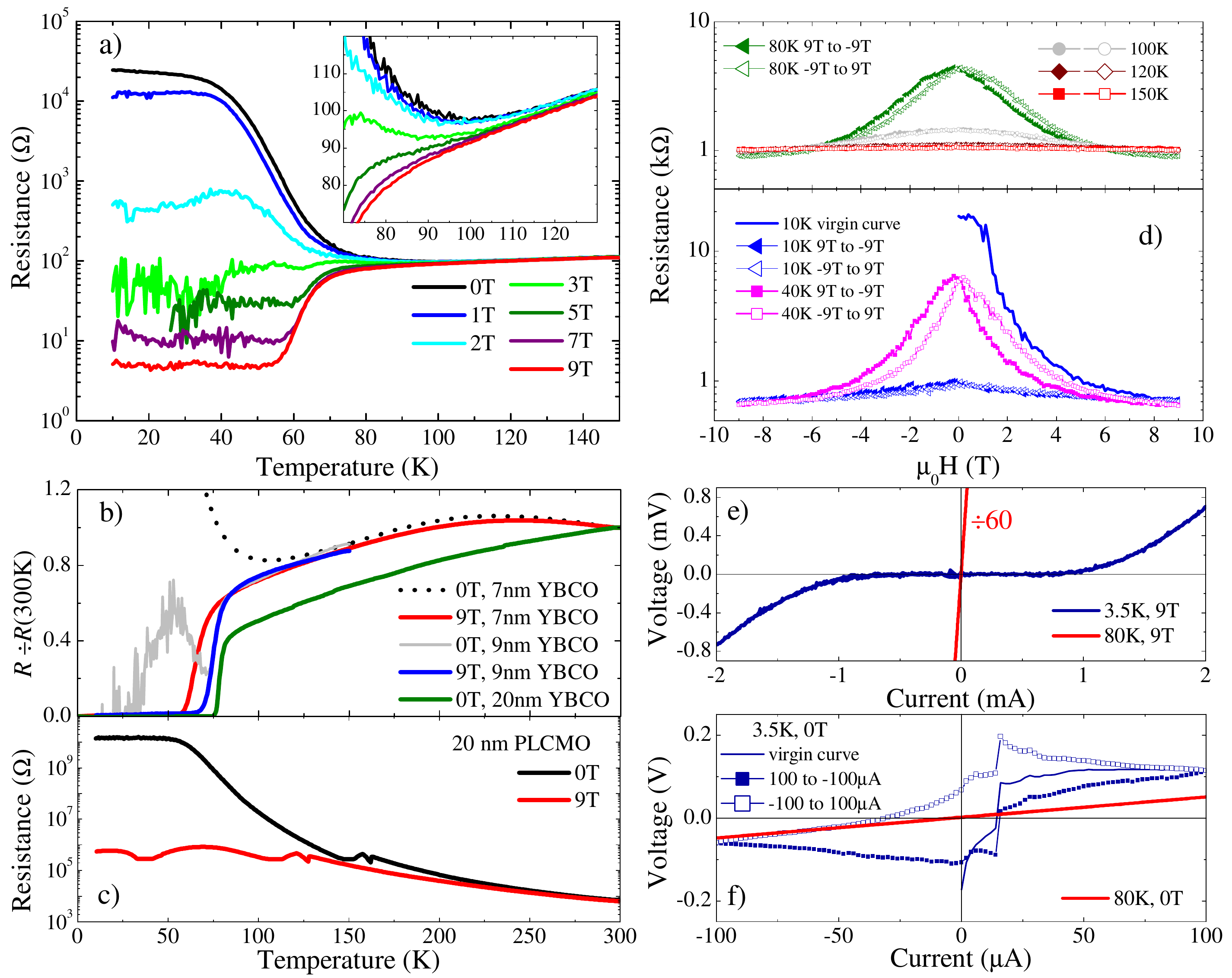}
	\caption{\label{fig1} 
	DC magneto-transport of PLCMO(20 nm)/YBCO($x$)/PCLMO(20 nm). (a) Resistance of a $x = 7$~nm sample measured whilst cooling in magnetic fields parallel to the layers. (b) Resistance normalized to the 300~K value for trilayers with $x$ as indicated in the legend. (c) Resistance of a single PLCMO film. (d) Magneto-resistance for $ x=7 $~nm. The applied current is 10~$\mu$A for (a)-(c) and 200~nA for (d). Voltage versus applied current for $ x =7 $~nm, cooled in 10~$ \mu $A and (e) at 9~Tesla and (f) in zero field. The data in (a) to (d) were measured in top contact geometry, the ones in (e) and (f) with a side contact geometry (see Fig.~4a of Ref.~\onlinecite{pypPRLsom}).}
\end{figure*} 

The magnetic-field-induced transition from a resistive to a coherent SC state is shown in Fig.~\ref{fig1} for a PLCMO(20 nm)/YBCO(7 nm)/PLCMO(20 nm) trilayer. Figures~\ref{fig1}a and b display the resistance versus temperature ($ R $-$ T $) curves measured with a top-contact geometry during cooling in different fields parallel to the layers. Similar results were obtained with perpendicular fields as shown in Fig. S4 of \cite{pypPRLsom}. Corresponding $ R $-$ T $ curves of a single PLCMO layer are displayed in Fig.~\ref{fig1}c. These data reveal nothing unusual for the normal state response of the YBCO layer which appears to be metallic. Between about 110 and 200~K, the $ R $-$T$ curves are indeed linear with moderate values of $R$. The broad maximum in the $R$-$T$ curve around 220~K and the decrease of $R$ toward 300~K (red line in Fig.~\ref{fig1}b) are probably due to additional conduction through PLCMO.  

Toward lower temperature, there is a drastic change in the response of the YBCO layer. In zero field the $R$-$ T $ curve exhibits a steep upturn at low $T$. In parallel, the magneto-resistance becomes very large and strongly hysteretic (see Fig.~\ref{fig1}d). Notably, the magnetic field leads to a strong reduction of $ R $ and at 3~Tesla the value of $R$ is already comparable to one above 100~K. There are also pronounced intrinsic fluctuation effects which lead to large jumps in the $R$-$ T $ curves. Increasing the field further inverts the shape of the $R$-$ T $ curve with a sharp decrease developing below about 70~K. At the highest field of 9~Tesla, this resistance drop is almost complete. An even larger field may be required to fully restore a phase coherent SC response. Alternatively, the small residual $R$ may be related to the top contact geometry due to an offset voltage from trapped charges in the upper PLCMO layer. 

Using a side-contact geometry we indeed found that the low-$ T $ resistance of YBCO at 9~Tesla vanishes (within our accuracy). The shape of the corresponding $ I $-$ V $ curve at 3.5~K and 9~Tesla in Fig.~\ref{fig1}e is characteristic of a coherent SC response. The voltage remains zero up to about 1~mA, above which it starts to increase. This yields an estimate of the critical current density of the 7~nm thick YBCO layer of $ j_{c}\approx 3\times 10^{3} $~A/cm$ ^{2} $. This value is still much smaller than for PLD-grown films of optimally doped YBCO with $ j_{c}\sim 10^{6}$-$10^{7} $~A/cm$ ^{2} $ \cite{hammerl2000}. Nevertheless, we have preliminary results (not shown) which indicate that $j_{c}$ increases strongly towards even larger magnetic fields. In the highly resistive state at low-$ T $ and zero-field, the $ I $-$ V $ curve shown in Fig.~\ref{fig1}f is strongly hysteretic and has an anomalous shape. The latter depends on the history of temperature, magnetic field and current. Notably, a large current tends to reduce the measured voltage. On the other hand, the $ I $-$ V $ curves at 80~K at zero and 9~Tesla are almost linear and ohmic (except for a small and strongly hysteretic offset voltage).

 \begin{figure*}

 		\includegraphics[width=1.0\textwidth]{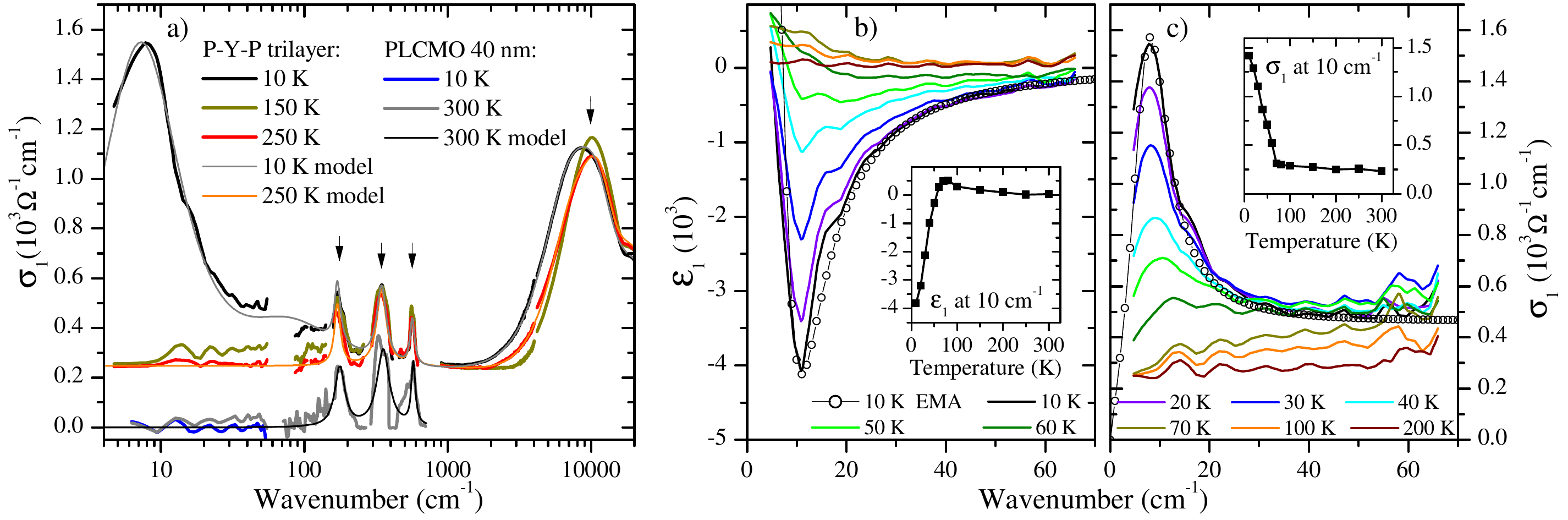}
 	\caption{\label{fig2}  
 	Zero-field optical response of a PLCMO(20 nm)/YBCO(7 nm)/PLCMO(20 nm) trilayer and PLCMO(40 nm) film. (a) Real part of the optical conductivity, $ \sigma_{1} $, at selected temperatures. Arrows indicate the IR-active phonons of PLCMO and its lowest $ d $-$ d $ interband transition. Thin lines indicate fits with a Drude-Lorentz model. (b) Real part of the THz dielectric function, $ \epsilon_{1} $ and (c), $ \sigma_{1} $ of the trilayer showing the plasmonic mode due to the confined or pinned SC condensate below 70~K. The respective insets detail the $ T $-dependence at 10~cm$ ^{-1} $. Open symbols show a fit with an effective medium model. The wave-like features in the THz spectra of PLCMO and the trilayer at $ T \geq 70 $~K are likely experimental artifacts  \cite{pypPRLsom}.}
 	
 \end{figure*} 
 
The resistance curves in Fig.~\ref{fig1}a show similarities with the ones of a granular SC \cite{gantmakher2010}. One signature is the nearly field-independent transition which either gives rise to a sharp increase, due to a localization of Cooper pairs below 3~Tesla, or a corresponding decrease in the $R$-$ T $ curve below about 70~K above 3~Tesla where macroscopic SC coherence develops. The gradual onset of this divergence below about 110~K (inset of Fig.~\ref{fig1}a) might be due to precursor SC correlations \cite{wang2006, dubroka2011}. Another example is the moderate $R$ value at 3~Tesla ($R_{\Box} \approx 100 $-$ 200 $~$ \Omega $), where the shape of the $R$-$ T $ curve changes over from an insulator-like to a SC behavior, that is well below the quantum limit of $ R_{Q}=h/2e=6.45 $~k$ \Omega $ for a Bose-glass of fully phase-incoherent Cooper-pairs \cite{fisher1990}.

Additional evidence for the confinement of the Cooper-pairs on a mesoscopic scale has been obtained from THz and infrared spectroscopy at zero-field. Figure~\ref{fig2}a shows the optical conductivity, $ \sigma_{1} $, due to the combined response of the PLCMO and YBCO layers. The main features at 250~K are a strong peak around $ 10,000 $~cm$ ^{-1} $ due to the lowest $ d $-$ d $ interband transition of PLCMO, three narrow modes at 175, 345, and 565~cm$ ^{-1} $ due to infrared-active phonon modes, and a finite conductivity at low-frequency that is assigned to the metallic response of the trilayer. Notably, below about 70~K there are pronounced changes in this low-frequency response (see Figs.~\ref{fig2}b and c). The downturn of $ \epsilon_{1} $ to large negative values signals a strong enhancement of the inductive response that is typical for a SC with a loss-free condensate. The main difference with respect to a bulk SC is that the spectral weight of the condensate in $ \sigma_{1} $ is contained in a narrow mode at about 7~cm$ ^{-1} $ instead of a delta function at the origin. An alternative interpretation of this THz mode in terms of the phason mode of a pinned charge density wave (CDW) in PLCMO \cite{nucara2008, kida2002} is excluded by our THz data on a single PLCMO layer as shown in Fig.~\ref{fig2}a.

The spectral weight of this THz mode, as obtained with a Drude-Lorentz model (thin lines in Fig.~\ref{fig2}a), amounts to a plasma frequency of the YBCO layer of $ \omega_{\mathrm{pl,SC}} \approx 2400 $~cm$ ^{-1} $. This value is quite typical for such a thin YBCO layer and, if it was not for the fact that the spectral weight is contained in a finite frequency mode, could be taken as evidence for a bulk-like SC response. The open symbols in Figs.~\ref{fig2}b and c show that the THz data can be described with a phenomenological effective medium model of SC grains that are separated by a dielectric layer. However, as outlined in Ref.~\onlinecite{pypPRLsom}, it yields an unrealistically small volume fraction of the dielectric. A model that explicitly accounts for the destructive interference and pair breaking effects at the SC grain boundaries thus may be required. As is mentioned towards the end, one may even need to consider an intertwined state of the SC and CDW orders \cite{berg2009}.

We emphasize that an explanation of the granular structure in terms of a structural inhomogeneity of the YBCO layer, e.g. due to an intergrowth of PLCMO, is unlikely. The analysis shown in Ref.~\onlinecite{pypPRLsom} testifies for the epitaxial growth and the structural and chemical quality of these trilayers. Furthermore, this magnetic-field-induced effect has been reproduced in several trilayers and it would be difficult to understand that a magnetic field would restore the SC coherence. 

 \begin{figure}

 		\includegraphics[width=0.9\columnwidth]{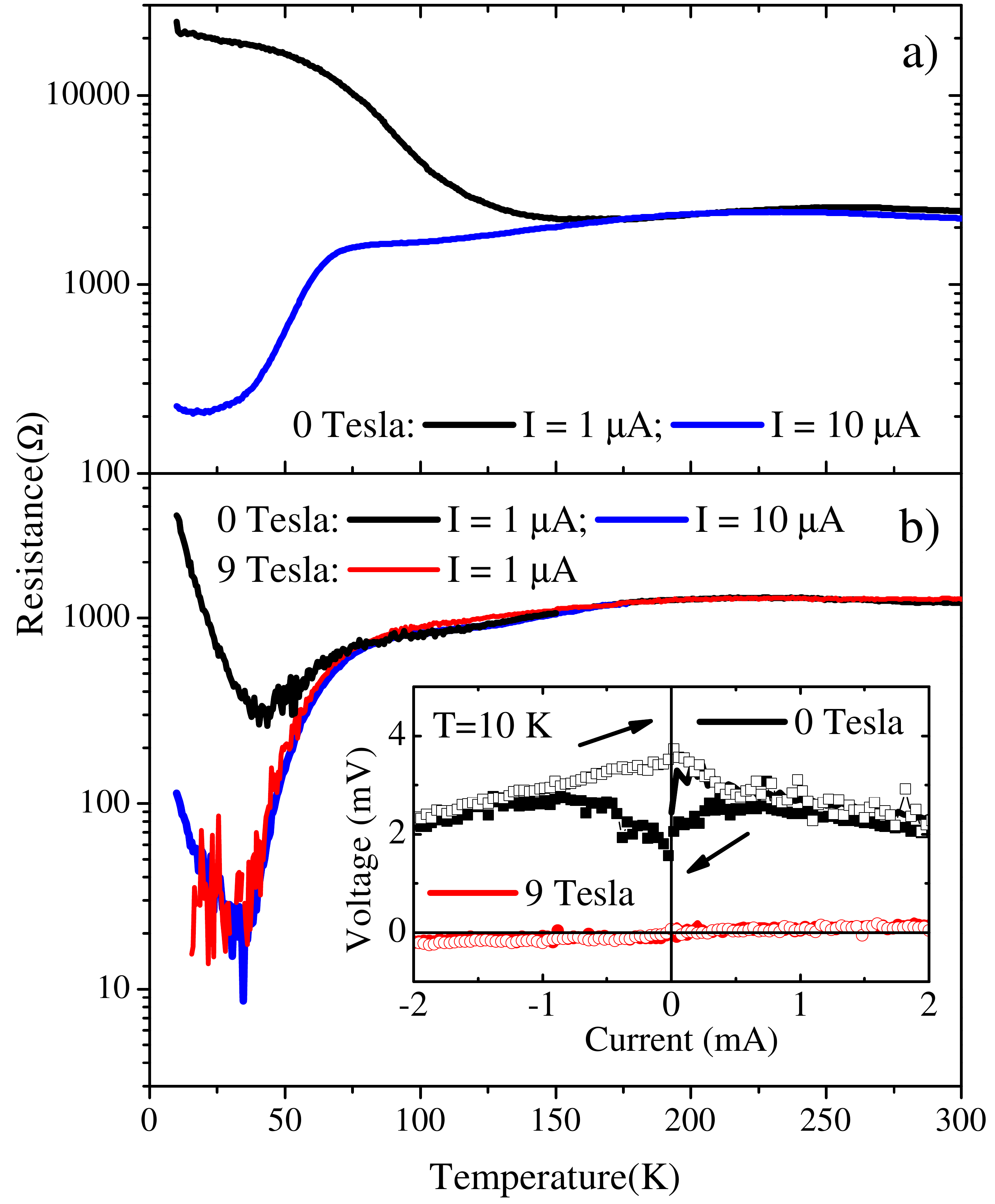}
 	\caption{\label{fig3}  
 	 $ R $-$ T $ curves for PLCMO (20 nm)/YBCO ($x$)/LaAlO$ _{3} $ (1 nm) bilayers with (a) $ x=3.5 $~nm and (b) $ x=4.5 $~nm. The inset to (b) shows the voltage vs. applied current at 10~K in 0 and 9~Tesla for $ x=4.5 $~nm after cooling in 1~$ \mu $A applied current and the labelled field. }
 	
 \end{figure} 

 \begin{figure*}
 	
 	\includegraphics[width=0.9\textwidth]{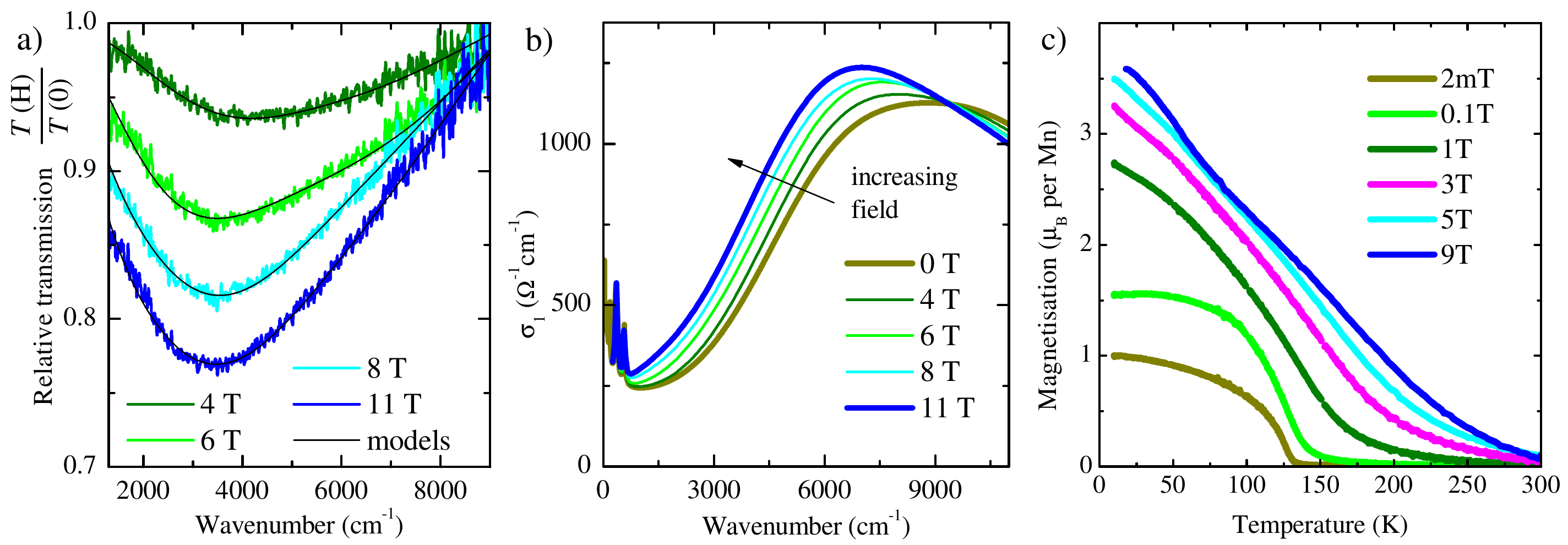}
 	\caption{\label{fig4} 
 		Effect of a magnetic-field parallel to the film on the charge/orbital order and magnetism of PLCMO in PLCMO(20 nm)/YBCO(7 nm)/PLCMO(20 nm).  (a) Optical transmission at $4.2$~K in selected fields normalized to the zero-field data. (b) Optical conductivity from modeling of the transmission and zero-field ellipsometry data (see Ref.~\onlinecite{pypPRLsom} for details). (c) Field-cooled magnetization. }
 	
 \end{figure*} 

Accordingly, we propose that this domain state in YBCO is induced by the interfacial coupling with PLCMO. The length scale of this interfacial effect appears to be about 4-5~nm from Fig.~\ref{fig1}b, which shows that the resistive upturn in zero field is already very weak for a trilayer with $ 9 $~nm of YBCO and entirely absent for 20~nm of YBCO. This interface effect is corroborated by the $R$-$ T $ curves in Figs.~\ref{fig3}a and b for PLCMO(20~nm)/YBCO($ x $) bilayers with $ x=3.5 $ and $ 4.5 $~nm, respectively. At 3.5~nm there is a strong resistive upturn towards low temperature at zero magnetic field. For $ x=4.5 $~nm this upturn is already very weak and the onset temperature is strongly reduced. 

For these bilayers it is also seen that a larger transport current tends to restore the SC coherence, similar to a large magnetic field. The bilayers allow us to directly probe the transport through YBCO, without the complications that may arise from currents through PLCMO and subsequent charge trapping (offset voltages) or current-induced conducting pathways~\cite{tokura1999,westhauser2006}. Yet, the $ I $-$ V $ curve at 10~K and zero Tesla of the $ x=4.5 $~nm sample in the inset to Fig.~\ref{fig3}b still shows a hysteretic behavior with an offset voltage at zero current that presumably arises from trapped charges at the boundaries between the SC grains. These charge freezing effects are also strongly suppressed by a large magnetic field. A more detailed understanding of the electronic response of this granular quantum state requires studies on mesoscopic samples with a better control of the conduction path with respect to the voltage contacts and is beyond the scope of this article.  

Instead, we turn to the question about the origin of this interface effect. A modification of the interfacial magnetic and electronic properties of YBCO was already observed in multilayers with FM La$ _{2/3} $Ca$ _{1/3} $MnO$ _{3} $~\cite{chakhalian2006,chakhalian2007,satapathy2012}. However, for a similar YBCO layer thickness, their SC response is coherent \cite{malik2012} and the effect of a magnetic field on SC is generally much weaker \cite{pena2005, liu2012}. 

This puts the focus on the specific properties of PLCMO. Bulk Pr$ _{0.7} $Ca$ _{0.3} $MnO$ _{3} $ (PCMO) is known for its charge and orbital order and related structural distortions that can be suppressed by a large magnetic field which induces an itinerant FM state \cite{tokura1999, kuwahara1995, zimmermann1999}. A corresponding magnetic field effect on the PLCMO layers is evident from the optical data in Figs.~\ref{fig4}a and b. They reveal a partial softening of an electronic mode that emerges from the lowest interband transition at 1.2~eV \cite{kovaleva2004, okimoto1998}. A similar, yet complete softening of this mode was observed in bulk PCMO in the context of the field-induced suppression of the charge/orbital order and the subsequent formation of an itinerant FM state \cite{okimoto1999}. The concomitant enhancement of the FM order is evident from the DC magnetization in Fig.~\ref{fig4}c. A diamagnetic response due to the SC grains is hardly visible here. However, as shown in Ref.~\onlinecite{pypPRLsom}, it can be seen for a perpendicular orientation of the magnetic field for which the SC screening currents are along the film plane (and the CuO$ _{2} $ planes of YBCO).

It remains to be understood whether the coupling mechanism between PLCMO and YBCO is of an electronic, magnetic or structural (possibly strain-related) origin. In the following we describe one possible scenario that a static CDW is induced in the YBCO due to the coupling to the charge-orbital order of the PLCMO. An incipient CDW, which is pinned by defects, has indeed been observed in underdoped YBCO crystals \cite{wu2011,ghiringhelli2012}. In the present case, an additional pinning mechanism in the YBCO layer may be provided by the Coulomb and strain fields that arise from the domain boundaries of the charge-orbital order and the related structural distortion in the adjacent PLCMO layers. Furthermore, the CDW and SC orders may be intertwined such that the SC order parameter acquires a finite wave vector and becomes spatially modulated \cite{berg2009}. Such a coupled CDW/SC domain state with strong pinning and pair-breaking effects at the domain boundaries and at defects, could indeed explain the granular nature of SC in the PLCMO/YBCO multilayers. Further studies, for example with resonant x-ray absorption and diffraction techniques of the Mn and Cu-specific magnetic and CDW orders will be required to elucidate the relevant coupling mechanism between the YBCO and PLCMO layers. Possibly, these studies will even help to further elucidate the intrinsic relationship between the CDW and SC in the bulk cuprates.

\begin{acknowledgments}
The work at UniFr has been supported by the Swiss National Foundation (SNF) through grants No. 200020-153660 and CRSII2-154410/1. 
\end{acknowledgments}


%

\end{document}